\documentclass[prd,twocolumn,nofootinbib,showpacs,floatfix,superscriptaddress]{revtex4}

\usepackage{amsmath}
\usepackage{graphicx}
\usepackage{bbm}
\usepackage{bm}
\usepackage{latexsym}
\usepackage{amssymb}
\usepackage{booktabs}
\usepackage[ansinew]{inputenc}
\usepackage{rotating}
\usepackage{euscript}
\usepackage{gensymb} 
\usepackage{slashed}
\usepackage{array}

\newcommand{\dd}{{\rm d}}

\begin{document}

\title{Threshold-Resummed Cross Section for the Drell-Yan Process \\[2mm] in 
Pion-Nucleon Collisions at COMPASS}

\author{Matthias Aicher}
\author{Andreas Sch\"afer}
\affiliation{Institute for Theoretical Physics, University of
Regensburg, D-93040 Regensburg, Germany}

\author{Werner Vogelsang}
\affiliation{Institute for Theoretical Physics,
                T\"{u}bingen University,
                Auf der Morgenstelle 14,
                D-72076 T\"{u}bingen, Germany}

\date{\today}

      \begin{abstract}
      We present a study of the Drell-Yan process in pion-proton 
collisions including next-to-leading-logarithmic threshold-resummed 
contributions. We analyze rapidity-integrated as well as rapidity-differential cross sections in the kinematic 
regime relevant for the COMPASS fixed target experiment. We find that 
resummation leads to a significant enhancement of the cross section 
compared to  fixed-order calculations in this regime. 
Particularly large corrections arise at large forward and backward rapidities 
of the lepton pair. We also study the scale dependence of the cross section 
and find it to be substantially reduced by threshold resummation.
       
      \end{abstract}
\pacs{12.38.Cy,12.38.Bx,14.40.Be}
\maketitle

\section{Introduction}

Drell-Yan lepton pair production by incident charged pions is still the only 
available hadronic process that allows for the extraction of the internal 
``partonic'' structure of the pion in the valence region. Unfortunately, 
experimental data for this process are rather scarce 
\cite{Conway:1989fs,Bordalo:1987cr}. Several analyses of the available 
data have been performed using leading order (LO) or next-to-leading order 
(NLO) partonic cross sections, with the
goal of extracting the parton distribution functions of the pion 
\cite{Sutton:1991ay,Gluck:1999xe,Wijesooriya:2005ir}. The resulting 
LO or NLO valence distributions showed a linear ($\sim (1-x)^1$) or 
slightly faster fall-off at high $x$, at odds with theoretical predictions 
based on perturbative QCD \cite{Farrar:1979aw},
and calculations using Dyson-Schwinger equations \cite{Hecht:2000xa}, 
which prefer a much softer behavior $\sim (1-x)^2$ at high $x$. For a 
review, see \cite{Holt:2010vj}.

In a recent analysis \cite{Aicher2010} we showed that a valence 
distribution with a fall-off $\sim (1-x)^{2\pm 0.1}$ is in fact well consistent 
with the Drell-Yan data, {\it if} large logarithmic contributions 
arising near the threshold for the partonic reaction are taken into 
account to all orders in perturbation theory. These so-called ``threshold 
logarithms'' strongly enhance the partonic cross section
near threshold, so that a softer valence distribution is sufficient to 
describe the data. We also found in Ref.~\cite{Aicher2010} that the 
available pionic Drell-Yan data are not able to completely 
determine the valence parton distribution of the pion. The data are 
almost equally well described by a valence 
distribution that carries 60 \%, 65 \% or 70 \% of the pion's momentum at 
a low input scale $Q_0 = 0.63$ GeV, as long as it behaves approximately 
as $(1-x)^2$ at high $x$. The quality of three corresponding fits
that we performed in \cite{Aicher2010} only differed by about one unit in 
$\chi^2$. This uncertainty in the valence momentum is also manifest in 
the earlier NLO analyses. At $Q = 2$ GeV the valence parton distribution 
of SMRS \cite{Sutton:1991ay} carries 46 \% of the pion's momentum, whereas 
the valence distribution of GRS \cite{Gluck:1999xe} carries only 40 \%, 
although both distributions describe the same data sets equally well. The 
origins of this ambiguity are the large overall systematic uncertainties 
of the Drell-Yan data. These uncertainties can be best discussed 
by introducing an additional $K$-factor that multiplies the theoretical 
cross section. The numerical value of this $K$-factor is strongly correlated  
with the first moment of the chosen valence distribution (see Table I 
of Ref.~\cite{Aicher2010}). Hence, pion Drell-Yan data with a well understood 
normalization are urgently needed to really pin down the pion's
valence distribution.

The upcoming fixed-target $\pi N$ Drell-Yan experiment~\cite{compass} 
at COMPASS is hoped to resolve this issue. In this paper,
we present detailed predictions for the rapidity-integrated as well as the 
rapidity-differential cross section for the kinematics relevant at 
COMPASS. In the light of our study~\cite{Aicher2010}, it is expected
that threshold logarithms will also play a significant role for 
the Drell-Yan cross section at COMPASS and lead to large corrections.
We will therefore base our predictions on the threshold resummation
technique. We note that there have been numerous earlier phenomenological
applications of threshold resummation in the Drell-Yan 
process~\cite{Shimizu:2005fp,Mukherjee:2006uu,Bolzoni:2006ky,Becher:2007ty,Ravindran:2006bu,Li:2009bra,Bonvini:2010tp}, both for fixed-target and
for collider energies. The specific application to
$\pi N$ scattering, and the resulting phenomenological studies for COMPASS,
are the new elements of this paper.

The remainder of this paper is organized as follows. In Sec.~\ref{sec2} 
the basic framework for the calculation of the rapidity-differential Drell-Yan 
cross section is presented. We discuss fixed-order corrections as well as 
the full next-to-leading logarithmic resummation of threshold logarithms. 
This includes the discussion of Mellin and Fourier moments of
the cross section, which are useful tools for threshold resummation.  
In Sec.~\ref{sec3} we present phenomenological results for the kinematic 
regime of the Drell-Yan experiment at COMPASS. We show both 
rapidity-integrated and rapidity-differential cross sections.
We finally draw our conclusions in Sec.~\ref{sec4}.

\section{Theoretical Framework \label{sec2}}
 
We follow in this section the framework laid out 
in Ref.~\cite{Mukherjee:2006uu}, where threshold resummation effects 
in $W^\pm$-boson production at hadron colliders were studied. 
We consider the inclusive cross section for the production of a 
$\mu^+\mu^-$ pair of invariant mass $Q$ and rapidity $\eta$ in the process 
\begin{equation}
 \pi^{-}(P_1) p(P_2) \to \mu^+  \mu^-  X,
\end{equation}
where $P_1$ and $P_2$ are the four-momenta of the initial-state particles. 
According to the relevant factorization theorem, at high $Q$ the 
rapidity-differential cross section may be written as
\begin{eqnarray}\label{eq:factorization}
\frac{\dd \sigma}{\dd Q^2 \dd \eta} &=& 
\sigma_0 \sum_{a,b} \int_{x_1^0}^1 \frac{\dd x_1}{x_1} \int_{x_2^0}^1 
\frac{\dd x_2}{x_2} f_a^{\pi} (x_1,\mu^2) f_b^p(x_2, \mu^2) \nonumber \\ 
& &\times 
e_{ab} \omega_{ab}(x_1,x_1^0,x_2,x_2^0,Q/\mu) ,
\end{eqnarray}
where $\sigma_0 = 4 \pi \alpha^2/9 Q^2 S$ with $S = (P_1 + P_2)^2$
the hadronic center-of-mass energy squared. 
The coupling $e_{ab}$ equals $e_q^2$ for the $q\bar{q}$ and 
$qg$, $\bar{q}g$ scattering processes which we are interested in,
where $e_q$ denotes the quark's fractional electromagnetic charge. 
In terms of the rapidity $\eta$
the lower bounds of the $x_1$ and $x_2$ integrals are
\begin{align}
 x_{1,2}^0 =\sqrt{\tau}\, e^{\pm \eta},
\end{align}
with $\tau= Q^2/S$. The sum in Eq.~(\ref{eq:factorization}) runs over all 
partonic channels, with $f_a^{\pi}$ and $f_b^p$ the corresponding parton 
distribution functions (PDFs) of the pion and the proton, and $\omega_{ab}$ 
the hard-scattering functions. The latter can be computed in perturbation 
theory as series in the strong coupling constant $\alpha_s$. The parton 
distribution functions as well as the hard-scattering functions depend on 
the factorization and renormalization scales, which we choose to be equal 
and collectively denote as $\mu$.

At leading order ${\cal O}(\alpha_s^0)$ only the quark-antiquark 
annihilation channel $q\bar{q}\to\gamma^*\to\mu^+\mu^-$ contributes, 
for which one has in our normalization
\begin{align}\label{eq:dy_lo}
\omega^{(0)}_{q\bar q} = x_1 x_2 \delta(x_1 - x_1^0) \delta (x_2 - x_2^0). 
\end{align}
At next-to-leading order, apart from the ${\cal O}(\alpha_s)$ 
corrections to the $q\bar{q}$ process, also additional processes contribute 
to the cross section, namely $qg\to\gamma^* q$ and $\bar q g\to 
\gamma^*\bar{q}$. The NLO partonic cross sections in the 
$\overline{\mathrm{MS}}$ scheme, which is the scheme we adopt
throughout this work, can be obtained from~\cite{Altarelli:1979ub}.
They are also collected in the Appendix of 
Ref.~\cite{Sutton:1991ay}.\footnote{We note that 
Refs.~\cite{Sutton:1991ay,Altarelli:1979ub}
adopt a non-standard polarization average for incoming gluons
in dimensional regularization. In order to correct for this, 
one simply needs to multiply the arguments of the logarithms in 
Eqs.~(A8) and~(A20) of \cite{Sutton:1991ay} by ${\mathrm{e}}^{-1}$ 
\cite{Gluck:1992tq}.}

As mentioned in the Introduction, the Drell-Yan cross section
receives large logarithmic corrections near the threshold for the
partonic reaction~\cite{Sterman:1986aj}. 
This threshold is defined by $z=Q^2/x_1 x_2S=1$, where 
$x_1$ and $x_2$ are the momentum fractions of the partons participating in the 
hard-scattering reaction. As $z$ increases towards unity, most of the initial 
partonic energy is used to produce the virtual photon. Therefore, little phase 
space remains for real-gluon radiation, while virtual-gluon diagrams 
may still contribute fully. The infrared cancellations between the
virtual and the ``inhibited'' real-emission diagrams then leave behind 
large logarithmic corrections to $\omega_{q\bar{q}}$. The
leading terms among the resulting threshold logarithms are of the 
form $\alpha_s^k\ln^{2k-1}(1-z)/(1-z)$ at the $k$th order of perturbation 
theory. Subleading terms are down by one or more powers of the 
logarithm. The threshold logarithms become particularly important
when $\tau = Q^2/S$, the hadronic analog of $z$, is large, which is generally 
the case in the fixed-target regime. The fact that the parton distribution 
functions are steeply falling functions of $x_1$ or $x_2$ emphasizes the 
threshold region in the cross section even for values of $\tau$ 
substantially smaller than one. In this kinematic regime the logarithms 
$\ln^{2k-1}(1-z)/(1-z)$ compensate the smallness of $\alpha_s^k$, and 
it becomes necessary to resum the large corrections to all 
orders in the strong coupling. Such ``threshold resummation'' has originally 
been derived for the Drell-Yan process and deep inelastic scattering a long 
time ago~\cite{Sterman:1986aj}. The techniques developed in these seminal 
papers have been extended and successfully applied to the resummation 
of large logarithmic contributions in numerous other hard QCD processes.

Threshold resummation may be achieved in Mellin moment space, where phase 
space integrals for multiple-soft-gluon emission decouple. For the rapidity 
dependent cross section, it is convenient to also apply a Fourier transform 
in $\eta$ \cite{Sterman:2000pt,Mukherjee:2006uu} (alternatively,
one can also use a double Mellin transform~\cite{Bozzi:2007pn}). 
Under combined Fourier and Mellin transforms of the cross section,
\begin{equation}
  \sigma(N,M) \equiv \int_0^1 \dd \tau \tau^{N-1} 
\int_{-\ln\frac{1}{\sqrt{\tau}}}^{\ln\frac{1}{\sqrt{\tau}}} 
\dd \eta e^{iM\eta} \frac{\dd \sigma}{\dd Q^2 \dd \eta},
\end{equation}
the convolution integrals in (\ref{eq:factorization}) 
decouple into ordinary products \cite{Mukherjee:2006uu, Sterman:2000pt}.
Defining the moments of the PDFs,
\begin{equation}\label{eq:mellin_moments}
 f^N(\mu^2) \equiv \int_0^1 \dd x x^{N-1} f(x,\mu^2),
\end{equation}
and introducing the corresponding double transform 
of the partonic hard-scattering cross sections,
\begin{equation}\label{eq:sigmaNM1}
 \tilde{\omega}_{ab}(N,M)\equiv
\int_0^1 \dd z z^{N-1} 
\int_{-\ln\frac{1}{\sqrt{z}}}^{\ln\frac{1}{\sqrt{z}}} 
\dd \hat\eta e^{iM\hat\eta} \omega_{ab},
\end{equation}
where $\hat\eta = \eta - \frac{1}{2}\ln(x_1/x_2)$ is the 
partonic center-of-mass rapidity, one finds:
\begin{equation}\label{eq:sigmaNM}
\sigma(N,M) = \sigma_0 \sum_{a,b} f_a^{\pi, N+i\frac{M}{2}}
f_b^{A, N - i \frac{M}{2}} e_{ab} \tilde{\omega}_{ab}(N,M).
\end{equation}
The double transform thus factorizes the PDFs and the perturbatively 
calculable hard-scattering functions. The leading order contribution to 
the hard-scattering function $\tilde{\omega}_{ab}(N,M)$ is easily calculated 
by making use of the relations
\begin{equation}
 \frac{x_1^0}{x_1} = \sqrt{z} e^{\hat\eta}, \quad \frac{x_2^0}{x_2} = 
\sqrt{z} e^{-\hat\eta}.
\end{equation}
We obtain from Eq. (\ref{eq:dy_lo})
for the Fourier transform of $\omega_{q\bar q}^{(0)}$:
\begin{equation}\label{eq:cosine}
\begin{aligned}
& \int_{-\ln(1/\sqrt{z})}^{\ln(1/\sqrt{z})} \dd \hat\eta e^{iM\hat\eta} 
x_1x_2 \delta(x_1 - x_1^0) \delta(x_2 - x_2^0) \\
& = \int_{-\ln(1/\sqrt{z})}^{\ln(1/\sqrt{z})} \dd \hat\eta e^{iM\hat\eta} 
\delta(1-\sqrt{z}e^{\hat\eta}) \delta(1-\sqrt{z}e^{-\hat\eta}) \\
& = \cos\left(M \ln(1/\sqrt{z})\right) \delta(1-z)\;.
\end{aligned}
\end{equation}
Here we have appropriately averaged over the two possible solutions for the 
integral. The emerging factor $\delta(1-z)$ in Eq. (\ref{eq:cosine}) 
is just the LO hard-scattering contribution to the rapidity-{\it integrated} 
Drell-Yan cross section. Hence, the Fourier transform of the LO 
rapidity-differential partonic cross section is equal to the LO 
rapidity-integrated partonic cross section times $\cos\left(M \ln(1/
\sqrt{z})\right)$~\cite{Mukherjee:2006uu}. In the near-threshold limit 
$z \to 1$ this cosine factor becomes subleading:
\begin{align}
 \cos\left(M \ln(1/\sqrt{z})\right) = 1 - \frac{(1-z)^2M^2}{8} + 
{\cal O}((1-z)^4 M^4).
\end{align}
As was discussed in Refs. \cite{Laenen:1992ey,Mukherjee:2006uu,Bolzoni:2006ky},
even at higher orders the dependence of the double moments 
$\tilde{\omega}_{ab=q\bar{q}}(N,M)$ on $M$ becomes subleading near threshold, 
whereas the $N$-dependence is identical to that of the rapidity-integrated 
cross section. Therefore, the resummed expression for 
$\tilde{\omega}_{ab=q\bar{q}}(N,M)$ is equal to that for the total 
(rapidity-integrated) cross section. It was shown in Ref. \cite{Mukherjee:2006uu} 
that keeping the cosine term in Eq.~(\ref{eq:cosine}) in the
resummed $\tilde{\omega}_{ab=q\bar{q}}$ slightly more faithfully 
reproduces the rapidity dependence at each order of perturbation theory.  
This can be easily achieved by writing the cosine as
\begin{equation}
  \cos\left(M \ln(1/\sqrt{z})\right) = \frac{1}{2}\left(z^{iM/2} + 
z^{-iM/2}\right),
\end{equation}
which, when combined with Eq.~(\ref{eq:sigmaNM1}), leads to a sum 
of two terms with Mellin moments shifted to $N \pm iM/2$.

Threshold resummation for the Drell-Yan process results in the 
exponentiation of the soft-gluon corrections. To 
next-to-leading-logarithmic (NLL) order the resummed 
cross section is given in the $\overline{\mathrm{MS}}$ scheme by 
\begin{eqnarray}\label{resummed}
 \ln\tilde{\omega}_{q\bar{q}}& =&
C_q\left(\frac{Q^2}{\mu^2},\alpha_s(\mu^2)\right) 
+ 2 \int_0^1 {\rm d}\zeta \frac{\zeta^{N-1} -1}{1-\zeta} \nonumber \\ 
&\times&\int_{\mu^2}^{(1-\zeta)^2 Q^2} \frac{{\rm d}k_\perp^2}{k_\perp^2} 
A_q(\alpha_s(k_\perp)),
\end{eqnarray}
where $A_q(\alpha_s)$ is a perturbative function, the $\mathcal{O}(\alpha_s^2)$ part of which is sufficient for resummation to NLL~\cite{Sterman:1986aj}:
\begin{equation}\label{exp_A}
 A_q(\alpha_s) = \frac{\alpha_s}{\pi} A_q^{(1)} + 
\left(\frac{\alpha_s}{\pi}\right)^2 A_q^{(2)}+ \dots,
\end{equation}
with \cite{Kodaira:1981nh}
\begin{equation}
 A_q^{(1)} = C_F, \quad A_q^{(2)} = \frac{1}{2} C_F\left[C_A
\left(\frac{67}{18}-\frac{\pi^2}{6}\right) - \frac{5}{9} N_f\right].
\end{equation}
Here $C_F=4/3$, $C_A=3$.
The first term in Eq.~(\ref{resummed}) does not originate from  
soft-gluon emission but instead mostly contains hard virtual 
corrections. It is also a perturbative series in $\alpha_s$, and 
we only need its first-order term:
\begin{equation}
C_q = \frac{\alpha_s}{\pi} C_F \left(-4 + \frac{2\pi^2}{3} + 
\frac{3}{2} \ln \frac{Q^2}{\mu^2}\right) +{\cal O}(\alpha_s^2),
\end{equation}
whose exponentiated form has been established in Ref. \cite{Eynck:2003fn}.

Since the perturbative running coupling $\alpha_s(k_\perp)$ diverges at 
$k_\perp = \Lambda_{\rm QCD}$, Eq.~(\ref{resummed}) as it stands is ill-defined. 
The perturbative expansion of the expression shows factorial divergence, 
which in QCD corresponds to a powerlike ambiguity of the series 
\cite{Beneke:2000kc}. It turns 
out, however, that the factorial divergence appears only at nonleading 
powers of the momentum transfer. The large logarithms we are resumming 
arise in the region \cite{Sterman:1986aj} $z \leq 1-1/\bar N$ in the 
integrand of the second term in Eq.~(\ref{resummed}). One therefore 
finds that to NLL they are contained in the simpler expression 
\begin{align}
 2\int_{Q^2/\bar N^2}^{Q^2}\frac{\dd k_\perp^2}{k_\perp^2} 
A_q(\alpha_s(k_\perp)) \ln \frac{\bar Nk_\perp}{Q} \nonumber \\ + 
2 \int_{Q^2}^{\mu^2} \frac{\dd k_\perp^2}{k_\perp^2} 
A_q(\alpha_s(k_\perp)) \ln \bar N\label{resu1}
\end{align}
for the second term in Eq.~(\ref{resummed}), where 
$\bar N = N e^{\gamma_E}$ with the Euler constant $\gamma_E$. 
This form is used for ``minimal'' expansions \cite{Catani:1996yz} 
of the resummed exponent. From Eq.~(\ref{resu1}) one obtains for 
the resummed exponent  to NLL accuracy \cite{Catani:1996yz,Catani:1998tm}:
\begin{equation}\label{eq:exponent}
\ln\tilde{\omega}_{q\bar{q}}=
C_q  + 2  h^{(1)}(\lambda) \ln \bar N 
+ 2 h^{(2)} \left(\lambda, \frac{Q^2}{\mu^2}\right), 
\end{equation}
where
\begin{equation}
\quad \lambda = b_0 \alpha_s(\mu^2) \ln \bar N.
\end{equation}
 The functions $h^{(1)}$, $h^{(2)}$ collect all leading-logarithmic
and NLL terms in the exponent, which are of the form 
$\alpha_s^k \ln^{k+1} \bar{N}$ 
and $\alpha_s^k \ln^{k} \bar{N}$, respectively. They read
\begin{eqnarray}\label{eq:h1}
h^{(1)}(\lambda) &=& \frac{A_q^{(1)}}{2 \pi b_0 \lambda}\left[2\lambda + 
(1- 2\lambda) \ln(1-2\lambda)\right],\nonumber \\
\label{eq:h2}
   h^{(2)}\left(\lambda, \frac{Q^2}{\mu^2}\right) &= & -
\frac{A_q^{(2)}}{2 \pi^2 b_0^2} \left[2\lambda + \ln (1-2\lambda)\right] 
\nonumber  \\
                   &  & +\frac{A_q^{(1)}b_1}{2\pi b_0^3}\left[2\lambda + 
\ln(1-2\lambda)\right. \nonumber  \\
& &\left.  + \frac{1}{2} \ln^2(1-2\lambda)\right] \nonumber  \\
                   &  & + \frac{A_q^{(1)}}{2\pi b_0} \left[2 \lambda + 
\ln(1-2\lambda)\right] \ln\frac{Q^2}{\mu^2} \nonumber \\
                   &  & - \frac{A_q^{(1)} \alpha_s(\mu^2)}{\pi} 
\ln (\bar N) \ln \frac{Q^2}{\mu^2},
\end{eqnarray}
where
\begin{eqnarray}
b_0 &=& \frac{1}{12\pi}\left(11 C_A - 2 N_f\right) \\
b_1 &=& \frac{1}{24 \pi^2}\left(17 C_A^2 - 5 C_A N_f -3 C_F N_f\right).
\end{eqnarray}
The last term of the function $h^{(2)}$ depends on the factorization 
scale and compensates the evolution of the parton distribution functions. 
The scale dependence of the second-to-last term results from the running 
of the strong coupling constant. Since scale evolution exponentiates and 
is therefore taken into account to all orders, one expects~\cite{scale,scale1} 
a significant decrease in the scale dependence of the resummed cross section 
compared to a fixed order cross section.

As was shown in Refs. \cite{Catani:2001cr,Kulesza:2002rh,Stavenga}, it 
is possible to improve the above formula by taking into account certain 
subleading terms in the resummation. As in Ref.~\cite{Mukherjee:2006uu} we rewrite 
Eqs.~(\ref{eq:exponent})-(\ref{eq:h2}) as
\begin{widetext}
\begin{equation}\label{eq:sigma_imp}
\begin{aligned}
\ln\tilde{\omega}_{q\bar{q}} = &\frac{1}{\pi b_0} \left[2 \lambda + 
\ln(1-2\lambda)\right] \left(\frac{A_q^{(1)}}{b_0 \alpha_s(\mu^2)} - 
\frac{A_q^{(2)}}{\pi b_0} +\frac{A_q^{(1)}b_1}{b_o^2} + A_q^{(1)} 
\ln \frac{Q^2}{\mu^2}\right) \\
 & + \frac{\alpha_s(\mu^2)}{\pi} C_F \left(-4 + \frac{2\pi^2}{3} \right) + 
\frac{A_q^{(1)} b_1}{2\pi b_0^3} \ln^2(1-2\lambda) + B_q^{(1)} 
\frac{\ln(1-2\lambda)}{\pi b_0} \\
 & + [-2A_q^{(1)} \ln \bar N - B_q^{(1)}] \left( \frac{\alpha(\mu^2)}{\pi} 
\ln \frac{Q^2}{\mu^2} + \frac{\ln( 1- 2\lambda)}{\pi b_0} \right),
\end{aligned} 
\end{equation}
\end{widetext}
where $B_q^{(1)} = -3 C_F/2$. The last term in Eq. (\ref{eq:sigma_imp}) 
is the leading-logarithmic expansion of the integral
\begin{equation}
\int_{\mu^2}^{Q^2/\bar N^2} \frac{\dd k_\perp^2}{k_\perp^2} 
\frac{ \alpha_s(k_\perp^2)}{\pi} [-2A_q^{(1)} \ln \bar N - B_q^{(1)}].
\end{equation}
The term in square brackets is the leading term in the large-$N$ limit 
of the anomalous dimension of the one-loop diagonal ($q \to q$) 
splitting function 
$P^N_{qq}$, i.e. it governs the evolution of the parton distributions 
between scales $\mu$ and $Q/\bar{N}$. Replacing it by the full flavor 
nonsinglet LO splitting function \cite{Kulesza:2002rh},
\begin{equation}
 [-2A_q^{(1)} \ln \bar N - B_q^{(1)}] \rightarrow C_F 
\left[\frac{3}{2}-2S_1(N)+\frac{1}{N(N+1)}\right],
\end{equation}
fully reproduces the diagonal part of the quark and anti-quark evolution.
This replacement further reduces the scale dependence of the resummed 
cross section. We could also include a non-diagonal contribution from 
$g\to q$ splitting, corresponding to singlet mixing. However, this
contribution turns out to be numerically unimportant for the Drell-Yan
process in the present kinematics.

As the exponentiation of soft-gluon corrections is achieved in Mellin 
moment and Fourier space, the hadronic cross section differential in 
$Q^2$ and $\eta$ is obtained by taking the inverse Mellin and Fourier 
transforms of 
Eq.~(\ref{eq:sigmaNM}):
\begin{equation}\label{eq:inverse}
   \frac{\dd \sigma}{\dd Q^2 \dd \eta} = \int_{-\infty}^{\infty} 
\frac{\dd M}{2\pi} e^{-iM\eta} \int_{C-i\infty}^{C+i\infty} 
\frac{\dd N}{2\pi i}\tau^{-N} \sigma(N,M).
\end{equation}
When performing the inverse Mellin transform, the parameter 
$C$ usually has to be chosen in such a way that all singularities 
of the integrand lie to the left of the integration contour. The 
resummed cross section, however, has a Landau singularity at $\lambda = 1/2$
or $\bar{N}=\exp(1/2\alpha_s b_0)$,
as a result of the divergence of the running coupling $\alpha_s$ 
in Eq.~(\ref{resummed}) for $k_\perp \to \Lambda_{\mathrm{QCD}}$. 
For the Mellin inversion, we adopt the \textit{minimal prescription} 
developed in Ref. \cite{Catani:1996yz} to deal with the Landau pole. For 
this prescription the contour is chosen to lie to the {\it left} of the Landau 
singularity. Above and below the real axis, the contour is tilted into 
the half-plane with negative real part. This improves the convergence
of the integration, since contributions with negative real part are 
exponentially suppressed by the factor $\tau^{-N}$ in Eq.~(\ref{eq:inverse}). 
As mentioned earlier, the moment-space singularities of the parton 
distribution functions are shifted by $\pm iM/2$ from the real axis due 
to the Fourier transform. Ref.~\cite{Sterman:2000pt} provides details for
how to prevent the tilted contour from passing through or below those 
singularities. We note that an alternative possibility for dealing with 
the Landau singularity is to perform the resummation directly in $z$-space 
\cite{Becher:2007ty}. 

We match the resummed cross section to 
the NLO one by subtracting the $O(\alpha_s)$ expansion of the resummed 
expression and adding the full NLO cross section \cite{Mukherjee:2006uu}. 
This ``matched'' cross section consequently not only resums the large 
threshold logarithms to all orders, but also contains the full NLO 
results for the $q\bar{q}$ and $qg$ channels. We will occasionally also 
consider a resummed cross section that has not been matched to 
the NLO one. We will refer to such a cross section as ``unmatched''.

\section{Phenomenological Results \label{sec3}}

We now present our numerical results for the Drell-Yan cross section at 
COMPASS. We will consider both the rapidity-integrated and the 
rapidity-differential hadronic cross section. Our main goal is to investigate
the size of the threshold resummation effects. The $\pi^-$  beam foreseen
at COMPASS has an energy of $190$ GeV. It is scattered off a proton target 
at rest, so that the resulting center-of-mass energy of the system is 
$\sqrt{S} \approx 19 \mbox{ GeV}$. 

For the pionic parton distribution functions we use the ones for 
the ``preferred fit'' of our previous study~\cite{Aicher2010}. 
We remind the reader that these were extracted from a fit to the earlier
$\pi N$ Drell-Yan 
data \cite{Conway:1989fs,Bordalo:1987cr}, using NLL threshold-resummed 
cross sections in the $\overline{\mathrm{MS}}$ scheme. For the proton 
target we use the NLO ($\overline{\mathrm{MS}}$ scheme) 
CTEQ6M \cite{Pumplin:2002vw} parton distributions. Unless stated 
otherwise, we choose the renormalization and factorization scales as 
$\mu = Q$. 

\begin{figure}[t]
  \includegraphics[height=\columnwidth, angle = 270]{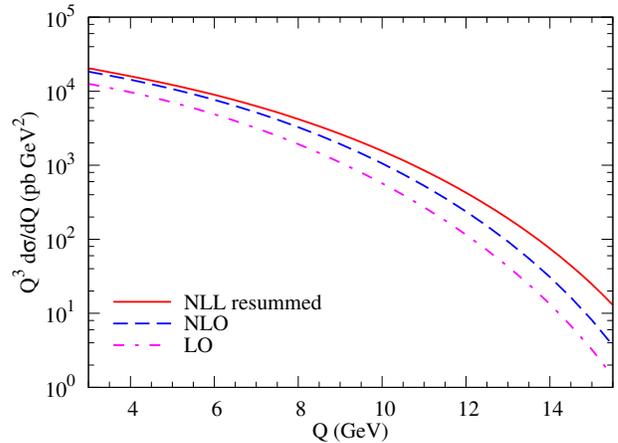}
  \caption{Rapidity-integrated Drell-Yan cross section $Q^3 \dd \sigma/\dd Q$ 
for $\pi^-$ $p$ scattering at $\sqrt{S} = 19 \mbox{ GeV}$, at LO, NLO and 
NLL-resummed, as a function of the invariant mass $Q$ of the lepton pair.}
  \label{fig:compass_int}
\end{figure}

\begin{figure}[t]
  \includegraphics[height=\columnwidth, angle = 270]{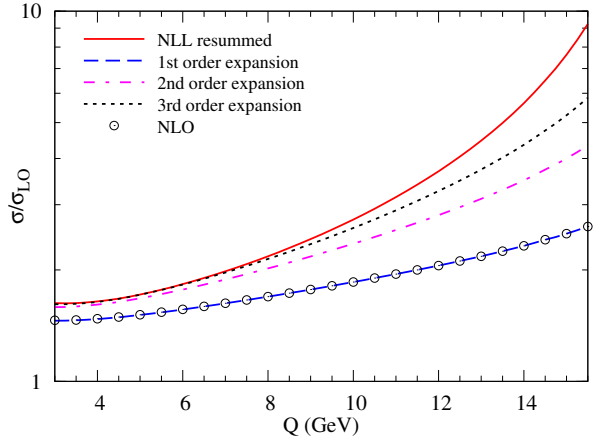}
  \caption{``K-factors'' as defined in Eq.~(\ref{eq:K-factor}) at 
$\sqrt{S} = 19 \mbox{ GeV}$ as functions of the lepton pair mass $Q$, 
at NLO (symbols) and for the NLL-resummed case. 
Also shown are the expansions of the resummed cross section 
to first, second and third order in the strong coupling.}
  \label{fig:compass_int_K}
\end{figure}

We start by considering the differential cross section $\dd \sigma / \dd Q$, 
integrated over all rapidities, to show the relevance and the validity 
of the resummation effects over the whole range of the invariant mass $Q$. 
Here we ignore for simplicity charmonium and bottonium resonances in 
the lepton pair spectrum, whose contributions are dominant for resonant 
invariant masses, and calculate only the smooth (continuum) part of the 
cross section. Figure~\ref{fig:compass_int} shows the cross section 
$Q^3 \dd \sigma / \dd Q$ at $\sqrt{S} = 19 \mbox{ GeV}$ at fixed order 
(LO and NLO), as well as for the NLL-resummed case. It can be seen 
that resummation leads to a significant enhancement of the cross
section over LO, which increases strongly with invariant mass. 
This becomes even more apparent in Fig. \ref{fig:compass_int_K}, where 
we show the ``K-factor'', defined as the ratio of the cross section 
to the LO one:
\begin{equation}\label{eq:K-factor}
K = \frac{\dd \sigma/\dd Q}{\dd \sigma^{{\mathrm{LO}}} / \dd Q}.
\end{equation}
The ``K-factor'' is plotted for the NLO and the NLL-resummed result. 
We also expand the {\it unmatched} resummed cross section in powers of 
$\alpha_s$. The results for the first, second and third order expansion 
are also shown in Fig. \ref{fig:compass_int_K}. One can see 
that in the fixed-target regime higher orders (beyond NLO) still 
make large contributions to the cross section, especially at high 
invariant mass $Q$. This finding is in line with that in the earlier 
study~\cite{Shimizu:2005fp} for $\bar{p}p$-scattering. We also observe
that the exact NLO cross section agrees extremely well with 
the first order expansion of the unmatched resummed result. This 
demonstrates that the logarithmic contributions from soft gluon radiation, 
which we resum to all orders, give by far the most important 
contribution to the cross section, not only very close to threshold as
$\tau = Q^2/S \to 1$, but also for rather moderate values of $\tau$.

\begin{figure}[t]
  \includegraphics[height=\columnwidth, angle = 270]{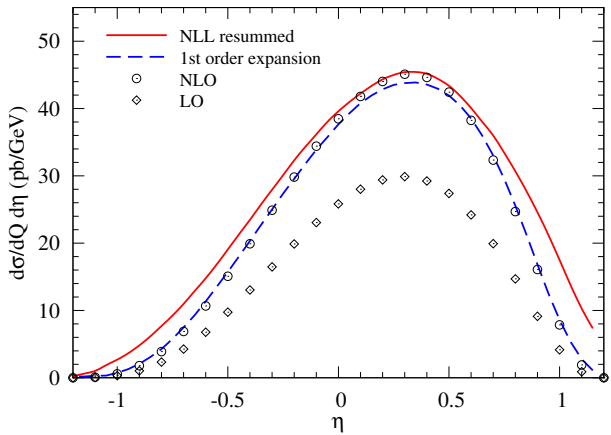}
  \caption{Rapidity-differential Drell-Yan cross section 
$\dd \sigma/\dd M \dd \eta$ for $\pi^-$ $p$ scattering at $\sqrt{S} = 
19 \mbox{ GeV}$ and $\sqrt{\tau}$ = 0.3. The LO, NLO and NLL-resummed 
cross sections as well as the first order expansion of the unmatched 
NLL-resummed cross section are shown as functions of the rapidity 
$\eta$ of the dimuon pair.}
  \label{fig:rap_03}
\end{figure}

\begin{figure}[t]
  \includegraphics[height=\columnwidth, angle = 270]{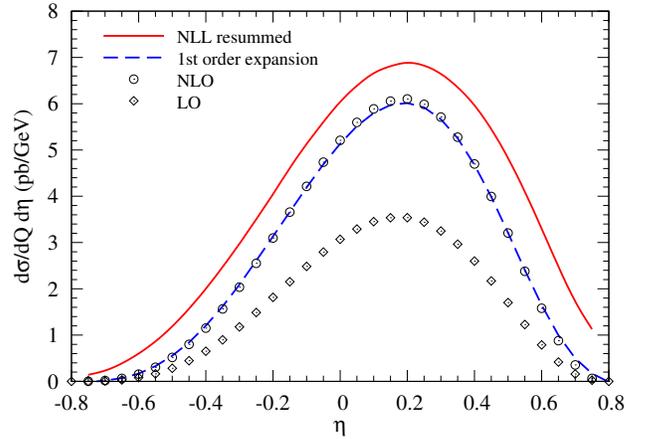}
  \caption{Same as Fig. \ref{fig:rap_03}, but at $\sqrt{\tau} = 0.45$.}
  \label{fig:rap_045}
\end{figure}

Next, we present the results for the rapidity distributions 
$\dd \sigma / \dd Q \dd \eta$. As mentioned above, charmonium and 
bottonium resonances complicate the calculation of Drell-Yan cross sections. 
Therefore usually only lepton pairs with invariant mass $Q$ between the 
$J/\Psi$ and $\Upsilon$ resonances and above the $\Upsilon$ 
are considered. Since the Drell-Yan event rate decreases rapidly with 
$\sqrt{\tau}$, it may not be possible to measure it above the $\Upsilon$ 
resonance in the medium-energy fixed-target regime accessed by the 
COMPASS experiment. We therefore make predictions for $\sqrt{\tau}  = 0.3$ 
and $\sqrt{\tau} = 0.45$, corresponding to $Q = 5.7$ GeV and $Q = 8.6$ GeV, 
respectively. Our results are presented in Figs. \ref{fig:rap_03} 
and \ref{fig:rap_045}. Again the resummed cross section and the 
fixed-order NLO and LO ones are shown. As before, we expand the 
unmatched resummed result in powers of $\alpha_s$ and find that the 
first order expansion agrees very well with the exact NLO result 
for $\sqrt{\tau} = 0.45$. For $\sqrt{\tau} = 0.3$, further away from
threshold, the first order expansion of the threshold resummed cross 
section lies very slightly below the exact NLO result for central rapidities. 
This is due to the fact that the contributions from the threshold region 
$z \to 1$ do not entirely dominate the cross section in this rapidity regime. 
As expected from our results in Figs. \ref{fig:compass_int} and 
\ref{fig:compass_int_K}, at fixed rapidity the threshold 
resummation effects become more important as $\tau$ increases, 
resulting in a fairly large enhancement of the resummed cross section 
at $\sqrt{\tau} = 0.45$. Nevertheless significant contributions from 
threshold resummation are still present in the cross section also 
for relatively modest values of $\tau$.

\begin{figure}[t]
  \includegraphics[height=\columnwidth, angle = 270]{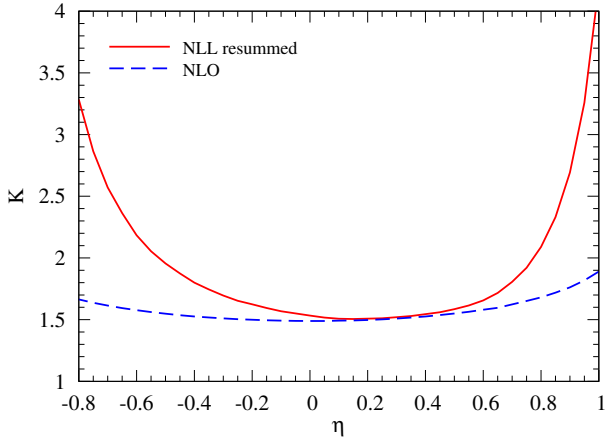}
  \caption{Ratios $K_{\rm res}$ and $K_{\rm NLO}$ as defined in 
Eq.~(\ref{eq:Kprime}) at $\sqrt{S} = 19 \mbox{ GeV}$ and $\sqrt{\tau} = 0.3$,
as functions of the rapidity $\eta$ of the dimuon pair.}
  \label{fig:rap_03_ratio}
\end{figure}

It is also interesting to examine to what extent resummation affects 
the {\it shape} of the rapidity dependent cross section. 
Figure~\ref{fig:rap_03_ratio} shows the ratios
\begin{equation} \label{eq:Kprime}
K_{\rm res} = 
\frac{\left({\displaystyle \frac{\dd \sigma_{\rm res}}
{\dd Q \dd \eta}}\right)}
{\left({\displaystyle\frac{\dd \sigma_{\rm LO}}
{\dd Q \dd \eta}}\right)}, 
\quad  K_{\rm NLO} = 
\frac{\left({\displaystyle \frac{\dd \sigma_{\rm NLO}}{\dd Q \dd \eta}}\right)}
{\left({\displaystyle \frac{\dd \sigma_{\rm LO}}{\dd Q \dd \eta}}\right)}
\end{equation}
as functions of the pair rapidity, at $\sqrt{S} = 19 \mbox{ GeV}$ and 
$\sqrt{\tau} = 0.3$. One can see that $K_{\rm res}$ becomes very large 
towards the boundaries of the $\eta$ interval. The resummed cross 
section shows a particularly sizable enhancement above the NLO one
at high rapidities. This enhancement is due to the fact that at 
fixed $\tau$ the limit $\eta \to \eta_{\rm max}$ corresponds to the 
limit $z \to 1$ at parton level. In this limit threshold logarithms 
become large regardless of the value of $\tau$. Large $\tau$ and/or 
high rapidities in
the fixed-target regime probe high momentum fractions $x$ in the 
parton distribution functions. Including threshold resummation in the 
analysis of parton distribution functions may hence have significant 
effects on their extracted high-$x$ behavior. This was examined 
recently in the context of the Drell-Yan process \cite{Aicher2010} and 
deep-inelastic lepton scattering~\cite{Corcella:2005us}.

\begin{figure}[t]
  \includegraphics[height=\columnwidth, angle = 270]{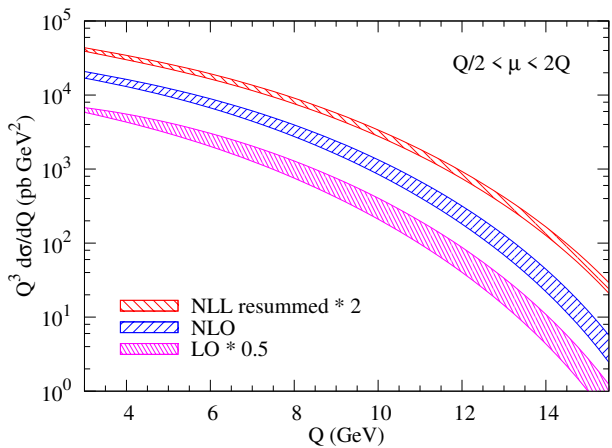}
  \caption{Scale dependence of the LO, NLO and NLL-resummed 
rapidity-integrated Drell-Yan cross sections at $\sqrt{S} = 19 \mbox{ GeV}$ as 
function of $Q$. The factorization as well as the renormalization 
scale have been varied between $Q/2$ and $2Q$. Note that we have 
multiplied the LO cross section by $1/2$ and the resummed cross 
section by $2$.}
  \label{fig:scales}
\end{figure}

\begin{figure}[t]
  \includegraphics[height=\columnwidth, angle = 270]{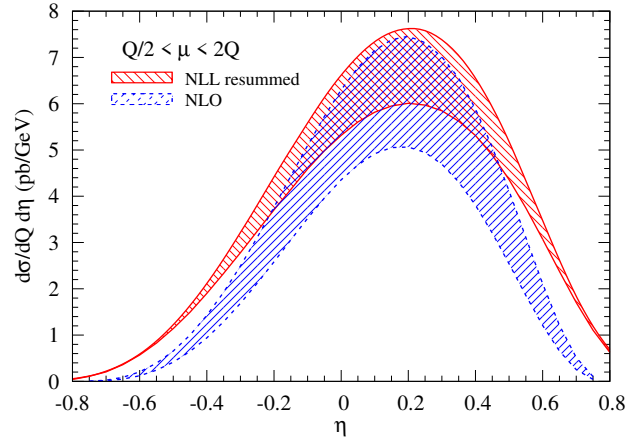}
  \caption{Scale dependence of the NLO and NLL-resummed rapidity-differential 
Drell-Yan cross sections at $\sqrt{S} = 19 \mbox{ GeV}$ and $\sqrt{\tau} = 
0.45$ as function of $\eta$. The factorization as well as the 
renormalization scale have been varied between $Q/2$ and $2Q$.}
  \label{fig:scales_rap}
\end{figure}

The crucial quality test for any higher order calculation is the extent to which it reduces the scale ambiguity inherent to any perturbative QCD calculation. We examine the scale dependences of the rapidity-integrated 
and the rapidity-differential cross sections in Figs. \ref{fig:scales}
and \ref{fig:scales_rap}, respectively. Again we show the LO, NLO and 
NLL-resummed results at $\sqrt{S} = 19 \mbox{ GeV}$, now varying
the renormalization and factorization scales between $\mu = Q/2$ 
and $\mu = 2Q$. Note that in Fig. \ref{fig:scales} we have 
for better visibility multiplied 
the LO cross section by $1/2$ and the resummed one by $2$. Evidently 
for the integrated cross section the scale dependence is decreased by
resummation over the whole range of invariant mass $Q$, whereas going 
from LO to NLO reduces the scale dependence only marginally. 
Figure~\ref{fig:scales_rap} shows the scale dependence of the 
rapidity distributions at $\sqrt{\tau} = 0.45$. Here we only show the 
NLL-resummed cross section and the NLO one. As one can see, the scale dependence is again 
significantly improved by resummation. This applies to all values
of rapidity; in fact the scale dependence almost vanishes at high $\eta$ 
after resummation.

\section{Conclusions \label{sec4}}

We have presented a phenomenological study of the Drell-Yan cross section 
for pion scattering off a proton target at COMPASS. In the calculation of 
the cross section we have resummed threshold corrections to next-to-leading 
logarithmic accuracy. The expansion of the resummed cross section to 
$\mathcal{O}(\alpha_s)$ agrees very well with the exact fixed-order calculation. 
This agreement demonstrates that the large threshold logarithms indeed 
dominate the Drell-Yan cross section and need to be taken into 
account to all orders. Resumming those logarithms leads to a significant 
enhancement above fixed-order calculations, even for moderate values of 
the invariant mass $Q$ of the lepton pair. We have also considered the 
rapidity dependence of the cross section. We find that even in cases 
where there is only a modest enhancement of the rapidity-integrated 
cross section by resummation, the shape of the rapidity-{\it differential} 
cross section is affected very strongly by resummation at sufficiently
large forward or backward rapidities. Finally, we have shown that the scale 
dependence of the perturbative cross section is substantially reduced when 
threshold resummed contributions are included.

Our results overall demonstrate that threshold resummation effects will be
important in the analysis of future COMPASS data. While we have only 
addressed the spin-averaged Drell-Yan cross section in this paper, 
we stress that threshold resummation effects are expected to be equally
relevant also for corresponding spin-dependent cross sections, even though
they may have a tendency to cancel in spin asymmetries. We also note
that in the light of our study cross sections and spin asymmetries at 
measured transverse momentum $q_\perp$ of the lepton pair, which will be a 
particular focus of the investigations at COMPASS, will require additional 
theoretical consideration. 

\section{Acknowledgments}
We thank Oleg Denisov for useful communications.
This work was supported by BMBF. M.A. was supported by a grant of 
the ``Bayerische Elitef\"orderung''.
W.V.'s work has been supported by the U.S. Department of Energy 
(contract number DE-AC02-98CH10886) and by the 
Alexander von Humboldt-Foundation.

\end{document}